\documentclass[letterpaper]{article} 
\usepackage{aaai25}  
\usepackage{times}  
\usepackage{helvet}  
\usepackage{courier}  
\usepackage[hyphens]{url}  
\usepackage{graphicx} 
\usepackage{natbib}  
\usepackage{caption} 
\usepackage{amsmath}
\usepackage{amsfonts}
\usepackage{algorithm}
\usepackage{algorithmic}

\usepackage{newfloat}
\usepackage{listings}
\DeclareCaptionStyle{ruled}{labelfont=normalfont,labelsep=colon,strut=off} 
\floatstyle{ruled}
\newfloat{listing}{tb}{lst}{}
\floatname{listing}{Listing}
\title{ChatHouseDiffusion: Prompt-Guided Generation and Editing of Floor Plans}
\author{
    Sizhong Qin\textsuperscript{\rm 1},
    Chengyu He\textsuperscript{\rm 1},
    Qiaoyun Chen\textsuperscript{\rm 1},
    Sen Yang\textsuperscript{\rm 1},
    Wenjie Liao\textsuperscript{\rm 2},
    Yi Gu\textsuperscript{\rm 1},
    Xinzheng Lu\textsuperscript{\rm 1}\thanks{Corresponding author}
}
\affiliations{
    \textsuperscript{\rm 1}Department of Civil Engineering, Tsinghua University\\
    \textsuperscript{\rm 2}Department of Civil Engineering, Southwest Jiaotong University\\

    \{qsz23, hcy23, cqy23, yang-s22, guy22\}@mails.tsinghua.edu.cn, liaowj@swjtu.edu.cn, luxz@tsinghua.edu.cn
}

\usepackage{bibentry}

\begin{document}

\maketitle

\begin{abstract}

The generation and editing of floor plans are critical in architectural planning, requiring a high degree of flexibility and efficiency. Existing methods demand extensive input information and lack the capability for interactive adaptation to user modifications. This paper introduces ChatHouseDiffusion, which leverages large language models (LLMs) to interpret natural language input, employs graphormer to encode topological relationships, and uses diffusion models to flexibly generate and edit floor plans. This approach allows iterative design adjustments based on user ideas, significantly enhancing design efficiency. Compared to existing models, ChatHouseDiffusion achieves higher Intersection over Union (IoU) scores, permitting precise, localized adjustments without the need for complete redesigns, thus offering greater practicality. Experiments demonstrate that our model not only strictly adheres to user specifications but also facilitates a more intuitive design process through its interactive capabilities.

\end{abstract}

%

\section{Background}
\label{section:Background}
Automatic floor plan generation technology is essential for enhancing design efficiency and reducing costs in architecture planning. Previous studies often relied on bubble diagrams to generate room layouts \citep{weberAutomatedFloorplanGeneration2022}, requiring users to have a clear idea of the final design to produce a reasonable outcome in one step. However, if users are dissatisfied with the initial result, they must reconstruct the bubble diagram to generate a new outcome, resulting in poor flexibility. Additionally, it is not possible to make local adjustments based on the initial result, making it difficult to fully utilize the outcomes of each generation step.

To enhance user design efficiency, \citet{leng-etal-2023-tell2design} proposed utilizing natural language for interaction. However, due to limitations in the training dataset, this method still requires users to provide comprehensive room layout information, making it unsuitable for flexible input forms. Moreover, it only supports one-step generation, preventing local adjustments to the initial results.

To address these issues, this study introduces a large language model (LLM) to parse user input, employs graphormer \citep{ying2021graphormer} to encode the topological relationships of rooms, and uses a diffusion model to predict floor plans. Furthermore, by replacing the attention map, precise local edits of the floor plan design are achieved. 

\begin{figure}[ht]
    \centering
    \includegraphics[width=\linewidth]{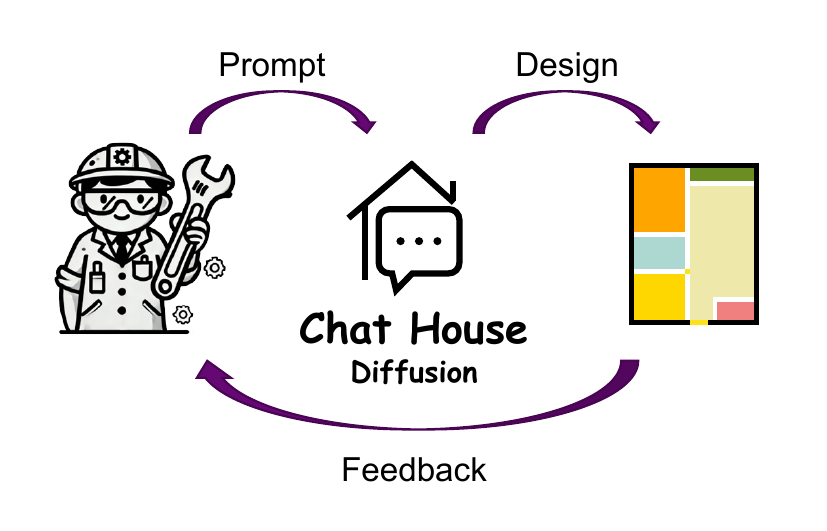}
    \caption{Iterative generation and editing of floor plans}
    \label{fig:concept}
\end{figure}

Based on the aforementioned method, a room layout design process that best meets practical design requirements can be constructed. Initially, users often have unclear design ideas. At this stage, they can input partial information to allow ChatHouseDiffusion to generate a preliminary floor plan and provide feedback to the users. Subsequently, users can iteratively adjust the floor plan, ultimately producing a satisfactory one after several iterations, as illustrated in Figure \ref{fig:concept}.

Section 2 introduces related work on room layout, transformer in graph, and editing in diffusion model. Section 3 provides a formal definition of the research problem. Section 4 describes the specific methods used in the study. Section 5 presents experiments and comparisons with previous methods. Section 6 provides the final conclusions.

\section{Related work}
\label{section: Related work}
\subsection{Floor plan generation}
\label{section:Floor plan generation}


Before the advent of deep learning algorithms, \citet{merrellComputergeneratedResidentialBuilding2010} utilized optimization algorithms to achieve room layout generation. With the development of deep learning, particularly generative AI technologies, the field of floor plan generation has seen a surge in significant research. \citet{wuDatadrivenInteriorPlan2019} pioneered the construction of the RPLAN dataset, laying the data foundation for subsequent studies.

HouseGAN \citep{nauataHouseGANRelationalGenerative2020b}, one of the early studies in this direction, was the first to use GANs to generate room layouts from topological relationships. HouseGAN++ \citep{nauataHouseGANGenerativeAdversarial2021a} improved upon HouseGAN’s model architecture and further achieved door generation. \citet{rahbarArchitecturalLayoutDesign2022} employed a two-stage generation method, first constructing bubble diagrams and then using cGAN to generate floor plans. \citet{luoFloorplanGANVectorResidential2022} optimized the feature representation of rooms, while \citet{upadhyayFloorGANGenerativeNetwork2023} proposed an end-to-end method that requires no post-processing. \citet{tangGraphTransformerGANs2023} introduced a Graph Transformer to express topological constraints, and \citet{aalaeiArchitecturalLayoutGeneration2023a} explored a vectorized approach to spatial layout representation.


In recent years, diffusion models have demonstrated superior capabilities compared to GANs, and several methods utilizing diffusion models for floor plan generation have emerged. HouseDiffusion \citep{shabaniHouseDiffusionVectorFloorplan2023} employs a diffusion model incorporating both discrete and continuous denoising processes to precisely control the generated floor plan structure, handling the floor plan’s room and door coordinates within a continuous coordinate system. \citet{suFloorPlanGraph2023} focuses on learning to generate reasonable floor plans in scenarios where topological knowledge is initially unknown. The study by \citet{guezeFloorPlanReconstruction2023} combines graph neural networks with constrained diffusion to reconstruct floor plans from sparse views, while \citet{zengResidentialFloorPlans2024} explores the automatic generation of residential floor plans under multiple conditional constraints using diffusion models.


In addition to GAN and diffusion models, some researchers have employed other deep learning methods. Graph2Plan \citep{huGraph2PlanLearningFloorplan2020} integrates Graph Neural Networks (GNN) and Convolutional Neural Networks (CNN) to process layout and floor plans, proposing a deep learning framework for generating floor plans from layout diagrams. \citet{paraGenerativeLayoutModeling2021} introduced a Transformer model to generate floor plans through constrained graph generation. WallPlan \citep{sunWallPlanSynthesizingFloorplans2022} couples a graph generation network GraphNet and semantics prediction network LabelNet to progressively generate wall diagrams by emulating graph traversal. \citet{duptyConstrainedLayoutGeneration2024} utilized a factor graph neural network to generate room layouts.

However, the aforementioned methods still rely on inputs like room topologies or bubble diagrams, resulting in limited interactivity. The integration of natural language processing has revolutionized the design process, making it more interactive and accessible. Tell2Design \citep{leng-etal-2023-tell2design} simplifies the process further by using deep learning to generate floor plans from natural language instructions, making it intuitive and user-friendly for individuals without expert knowledge in architectural design. ChatDesign \citep{liChatdesignBootstrappingGenerative2024} employs large pre-trained language models to convert textual descriptions into architectural designs, facilitating an iterative refinement process.

\subsection{Transformer for graph}
For graph structures, Transformers can utilize self-attention mechanisms to compute the semantic similarity between each individual node and other nodes in the graph, establishing connections between individual units and the overall structure. However, for graphs, it is necessary for models to perceive both the spatial structure of the graph and the relationships between nodes.

Graphormer \citep{ying2021graphormer} adopts three simple and efficient spatial encoding methods to leverage graph information, including Centrality encoding, spatial encoding, and edge encoding. GraphGPS \citep{rampasek2022GPS} provides a three-part approach to building graph Transformers with linear complexity, which involves positional/structural encoding, local message-passing, and global attention. \citet{Chen22aSAT} introduce a flexible structurally aware self-attention mechanism (SAT), which incorporates local substructures around each node when computing self-attention scores. This novel self-attention mechanism considers not only the similarity of node attributes but also the structural similarity between subgraphs.

\subsection{Training and finetuning free editing in diffusion models}
Due to the lack of necessary datasets for editing, researchers are focusing on implementing training and fine-tuning-free editing methods directly. In their comprehensive review, \citet{huang2024diffusion} identify five different implementation approaches, among which the attention modification method stands out as the most prevalent and straightforward for training-free image editing. A pioneering study in this field, Prompt2prompt \citep{hertzPrompttoPromptImageEditing2022}, achieves image editing by directly manipulating the attention map, replacing it to guide the editing process. Building on this, Pix2Pix-Zero \citep{Parmarpix2pix-zero} advances the technique by eliminating the need for user-defined text prompts in real image editing. This method autonomously discovers editing directions within the text embedding space while meticulously preserving the original content structure.

Further innovations include MasaCtrl \citep{cao_2023_masactrl}, which leverages mutual self-attention to modify Key and Value features for action editing. Similarly, PnP \citep{ju2023direct} utilizes spatial feature manipulation alongside self-attention to introduce guidance image features, focusing specifically on the Query and Key components.

On another front, Local Attention Map Modification techniques, such as TF-ICON \citep{lu2023tf} and Object-Shape Variation \citep{patashnik2023localizing}, fine-tune local attention maps for enhanced image composition and shape variation within text-to-image workflows. TF-ICON seamlessly integrates user-provided objects without additional training, while Object-Shape Variation employs prompt-mixing to offer diverse shape choices.

Lastly, Attention Score Guidance methods like Conditional Score Guidance \citep{Lee2023ConditionalScoreGuidance} and EBMs \citep{park2024energy} utilize attention score functions for selective image editing. These methods enhance semantic alignment and high-fidelity translation by providing adaptive context control, marking a significant advancement in the field of image editing.
\begin{figure*}[ht]
    \centering
    \includegraphics[width=\linewidth]{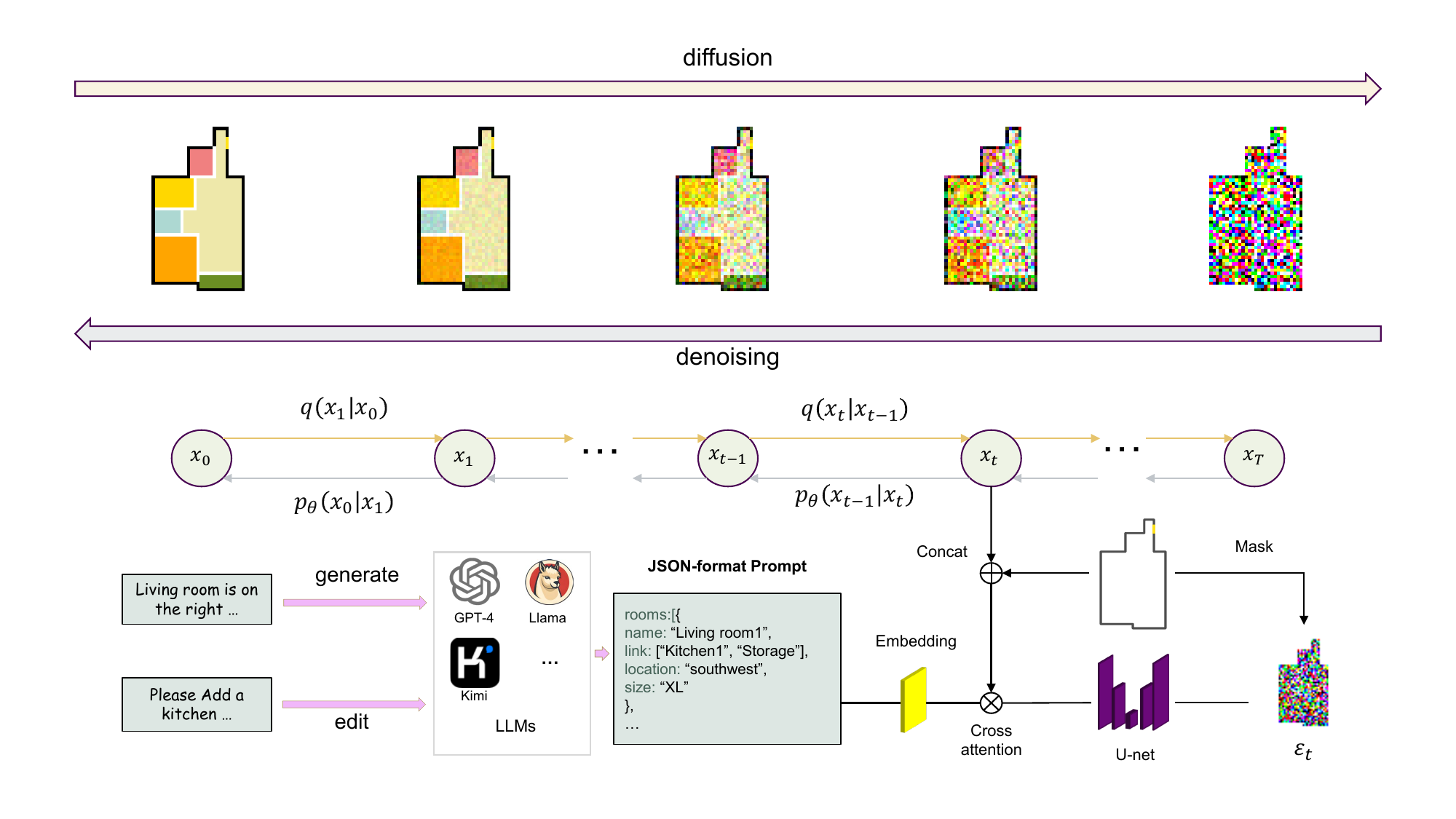}
    \caption{Framework of text prompt room plan generation based on diffusion models}
    \label{fig:framework}
\end{figure*}

\section{Formulation}
\label{section:Formulation}

The training problem for Denoising Diffusion Probabilistic Models (DDPM) for floor plan design can be mathematically formalized as the optimization problem:
\begin{align*}
    & \min_{\theta} \sum_{i=1}^{N} \mathbb{E}_{\mathbf{T}^{(i)}, \mathbf{I}^{(i)}} [-\log P(\mathbf{P}^{(i)} | \mathbf{T}^{(i)}, \mathbf{I}^{(i)}; \theta)] \\
    \iff & \min_{\theta}  \mathbb{E}_{\mathbf{T}^{(i)},\mathbf{I}^{(i)},x^{(i)}_0}\mathbb{E}_{\epsilon,t\sim[1,T]}[\Vert\epsilon-\epsilon_\theta(x_t,t,\mathbf{T}^{(i)},\mathbf{I}^{(i)})\Vert^2]
\end{align*}

where
    $\theta$ represents the parameters of DDPM,
    $N$ is the number of samples in the dataset $D$,
    $\mathbf{P}^{(i)}$ is the ground truth floor plan for the $i$-th sample,
    $P(\mathbf{P}^{(i)} | \mathbf{T}^{(i)}, \mathbf{I}^{(i)}; \theta)$ is the conditional probability of generating the ground truth floor plan $\mathbf{P}^{(i)}$ given the text prompt $\mathbf{T}^{(i)}$, the image $\mathbf{I}^{(i)}$, and the model parameters $\theta$. The optimization aims to minimize the negative log-likelihood of generating the ground truth floor plans conditioned on the provided text prompts and images over the entire dataset. This task involves predicting noise $\epsilon$ in a diffusion process where $t$ is the current step, $T$ the total steps, $x_t$ the noisy image at step $t$, and $\epsilon_\theta(x_t,t,\mathbf{T}^{(i)},\mathbf{I}^{(i)})$ the model's noise prediction.

\section{Method}
\label{section:Method}
This research aims to automate the generation and editing of floor plans using textual prompts and outlines, aiding designers in layout planning. Diffusion models are selected for their ability to produce varied and unique floor plans while avoiding mode collapse, ensuring stable and precise outputs. By integrating textual prompts with structured multi-modal inputs, the model's ability to understand and meet specific design requirements is enhanced, offering greater flexibility and control.  

\subsection{Framework}
\label{section:Framework}
\begin{figure*}[ht]
    \centering
    \includegraphics[width=\linewidth]{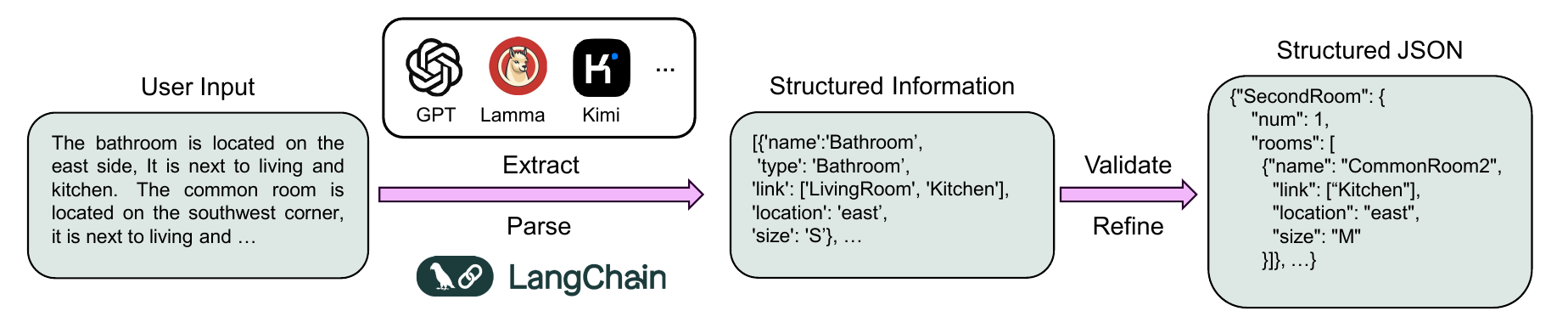}
    \caption{Workflow of Prompt to JSON}
    \label{fig:workflow}
\end{figure*}
The framework is displayed in Figure \ref{fig:framework}. It can be divided into two main parts: the interpretation of room design conditions in natural language form based on LLM, and the conditional generation or editing of floor plans based on the diffusion model. Furthermore, during the model training process, the steps of LLM parsing are substituted with manual rule-based parsing to ensure the accuracy of the training dataset.

For LLM interpretation, after receiving user input prompts, the LLM analyzes the text to obtain JSON-formatted prompts, which are then processed and used as generation conditions for the subsequent diffusion model.

To achieve the processes of diffusion and denoising, a neural network is constructed based on textual prompts and room outlines to predict noise. The room outline is firstly utilized as a mask to process the image $x_t$ at the $t$-th step. The outline and $x_t$ are then concatenated, and their cross-attention with the textual prompt's embedding is used as input features. Finally, a U-net is employed for prediction, and the room outline mask is applied to the final noise prediction. This method allows for the full utilization of room outline information during the generation and editing processes, ensuring that the diffusion model focuses on generating the floor plan within the outline, thereby avoiding other influences.

In the subsequent sections, we will delve into the specific methods for prompt-to-JSON conversion, plan generation, and plan editing.

\subsection{Prompt to JSON}
\label{section:Prompt to JSON}
The process of converting a user's natural language description into structured JSON is illustrated in Figure \ref{fig:workflow}. This process starts with the user's input, which includes a natural language description detailing room locations and relationships within a building.

The first major step involves extracting information using a LLM. Advanced AI models such as GPT, LLaMA, and Kimi are utilized at this stage. By constructing appropriate prompts, these models are enabled to recognize key details such as room names, types, locations, and their relationships. The objective here is to transform the unstructured input into structured data. 

Following extraction, the LangChain Parsing of the LLM output occurs, which is a tool used to parse the output into a structured format. The output is structured as a list of dictionaries, with each dictionary representing a room and its attributes such as name, type, location, and linked rooms.

The next phase involves enumeration type checks to verify the legality of the JSON attribute values. This step ensures that each attribute in the structured data adheres to predefined types and constraints. Additionally, fuzzy string matching based on Levenshtein Distance is employed to adjust JSON attribute values to the specified terms. This technique corrects minor discrepancies and standardizes attribute values. For instance, if a user inputs "bathrm" instead of "bathroom," fuzzy string matching would recognize and correct the term. Once the data is validated and refined, it is converted into a JSON-format prompt, making it ready for use in floor plan generation.

\subsection{Generation}
\label{section:Plan generation}
Upon obtaining the JSON-formatted text prompts, it is desired to represent the attributes and topological relationships of rooms in a more rational manner. Previous text-to-image methods \citep{saharia2022imagen} typically employ models such as T5 \citep{2020t5} to directly encode the text. While suitable for natural language, previous approach struggles to capture the topological relationships of different rooms.

\begin{figure}[ht]
    \centering
    \includegraphics[width=\linewidth]{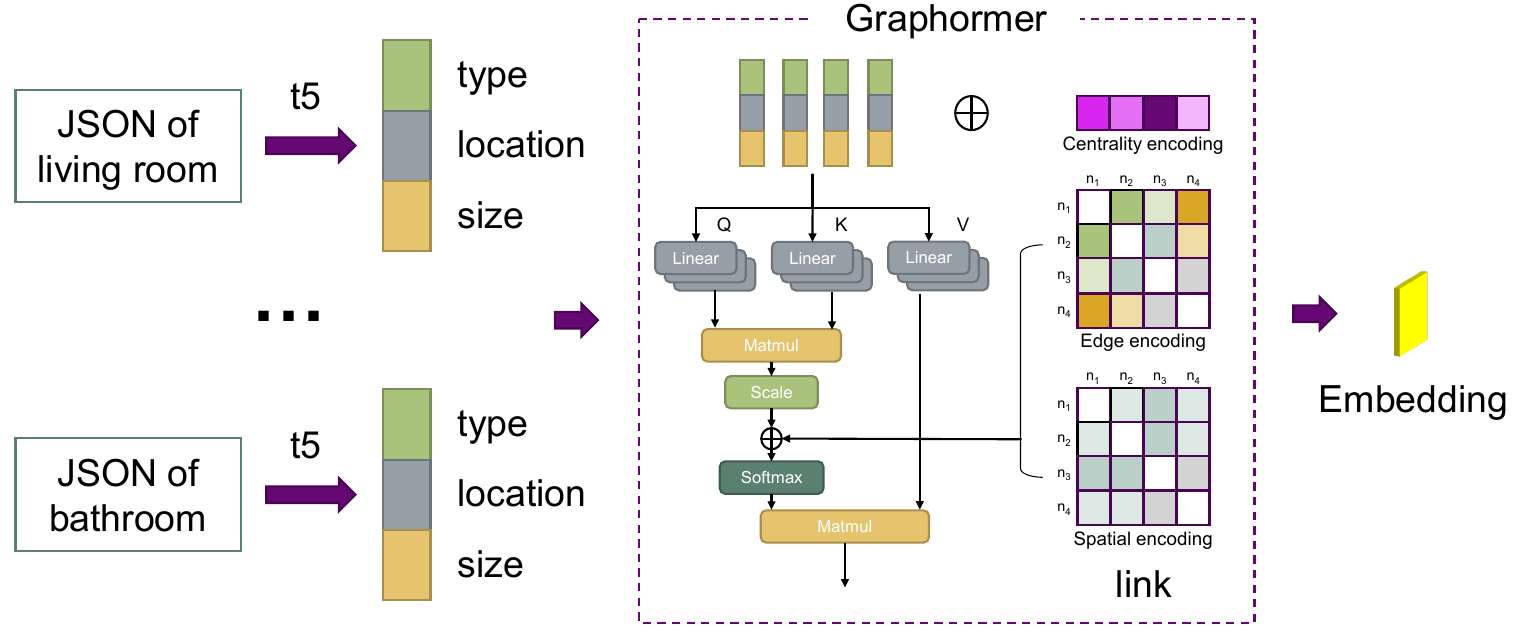}
    \caption{Floor plan information encoding method combining attribute features and topological relationships}
    \label{fig:generation}
\end{figure}

This study proposes a feature representation method combining the T5 encoder \citep{2020t5} and graphormer \citep{ying2021graphormer}, as illustrated in Figure \ref{fig:generation}. Initially, the type, location, and size information of each room are extracted from the structured room data. These details are encoded using a pre-trained T5 encoder and concatenated, converting each room into a feature vector. To further incorporate the topological relationships between rooms, the graphormer method is utilized. 

This method integrates the connections between different rooms with the Attention map, enabling the learning of topological information. Three types of encoding were constructed. Centrality encoding uses the degree to represent the importance of each room, which is then added to the input feature vector. Since the room layout is an undirected graph, the in-degree and out-degree are equal. Spatial encoding employs the unweighted shortest path to establish the importance of relationships between nodes. Edge encoding represents the information of edges. Since only the topological relationship between rooms is considered, all edge information is temporarily assigned a value of 1. Additional information, such as doors, can be introduced as needed in future work. The resulting output serves as the text embedding of the diffusion model.

\subsection{Editing}
\label{section:Plan editing}
\begin{figure}[ht]
    \centering
    \includegraphics[width=\linewidth]{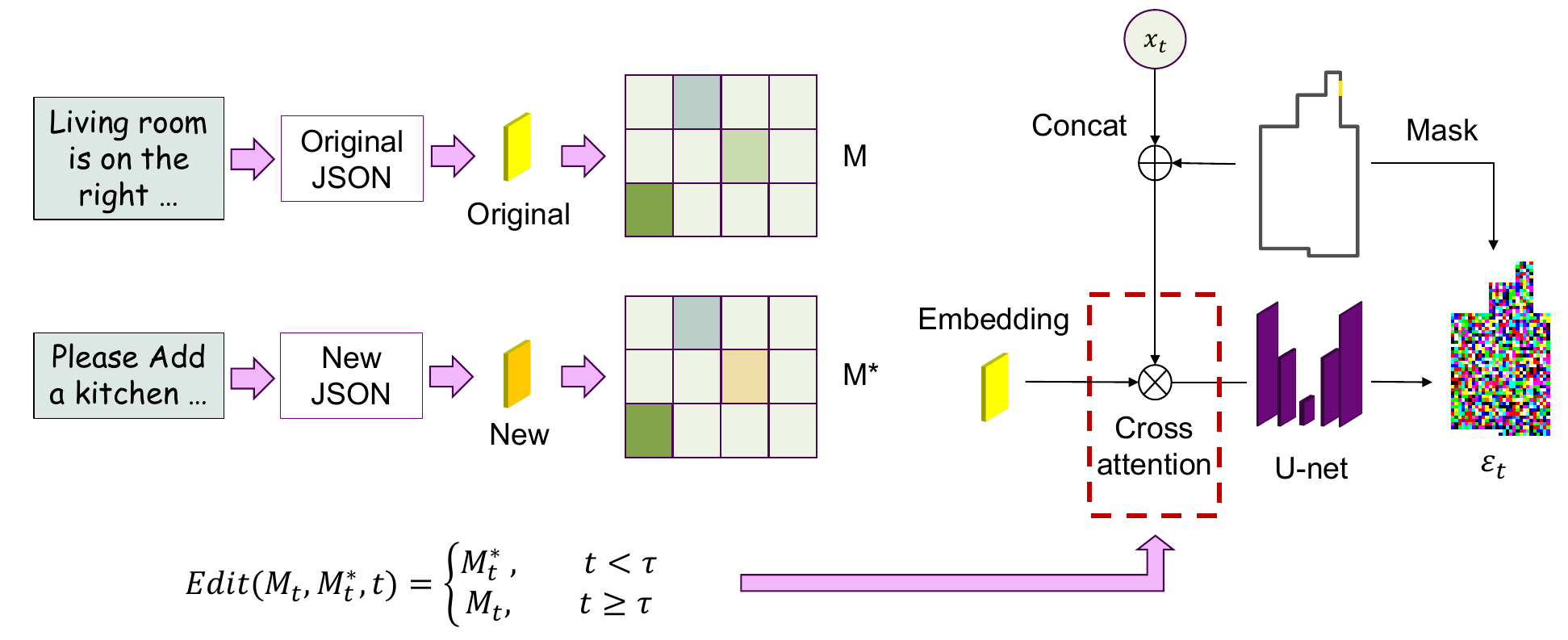}
    \caption{Floor plan editing method based on cross-attention map}
    \label{fig:editing}
\end{figure}

After generating the initial room plan, users may find certain localized results unsatisfactory but prefer to avoid global changes by regenerating the entire layout. Therefore, this section implements a localized editing method for the floor plan, as shown in Figure \ref{fig:editing}. 

This method is inspired by the Prompt to Prompt \citep{hertzPrompttoPromptImageEditing2022} technique, wherein the cross-attention map during the denoising process is replaced to maintain the overall layout information while modifying only the local details. 

During the editing process, a random seed is initially fixed, and all cross-attention maps from each step of the initial generation process are saved. Using the same random seed, new text embeddings are obtained through a large language model and graphormer. In the initial denoising steps, the original cross-attention maps replace the new cross-attention maps. The intensity of editing can be controlled by adjusting the threshold $\mathrm{\tau}$, thereby achieving precise floor plan editing.

\section{Experiment}
The computing platform specifications were as follows: OS: Ubuntu 22.04 LTS; CPU: Intel Xeon E5-2682 v4 @ 64 × 3 GHz; RAM: 32 GB; GPU: NVIDIA GeForce RTX 3090 24 GB.
\label{section:Experiment}
\subsection{Data}
\label{section:Data}
During the training phase, the RPLAN dataset \citep{wuDatadrivenInteriorPlan2019} is utilized, which is a manually collected large-scale densely annotated dataset of floor plans from real residential buildings. Due to our task being very similar to existing studies \citep{leng-etal-2023-tell2design}, we have taken their result as the baseline.

\begin{figure}[ht]
    \centering
    \includegraphics[width=\linewidth]{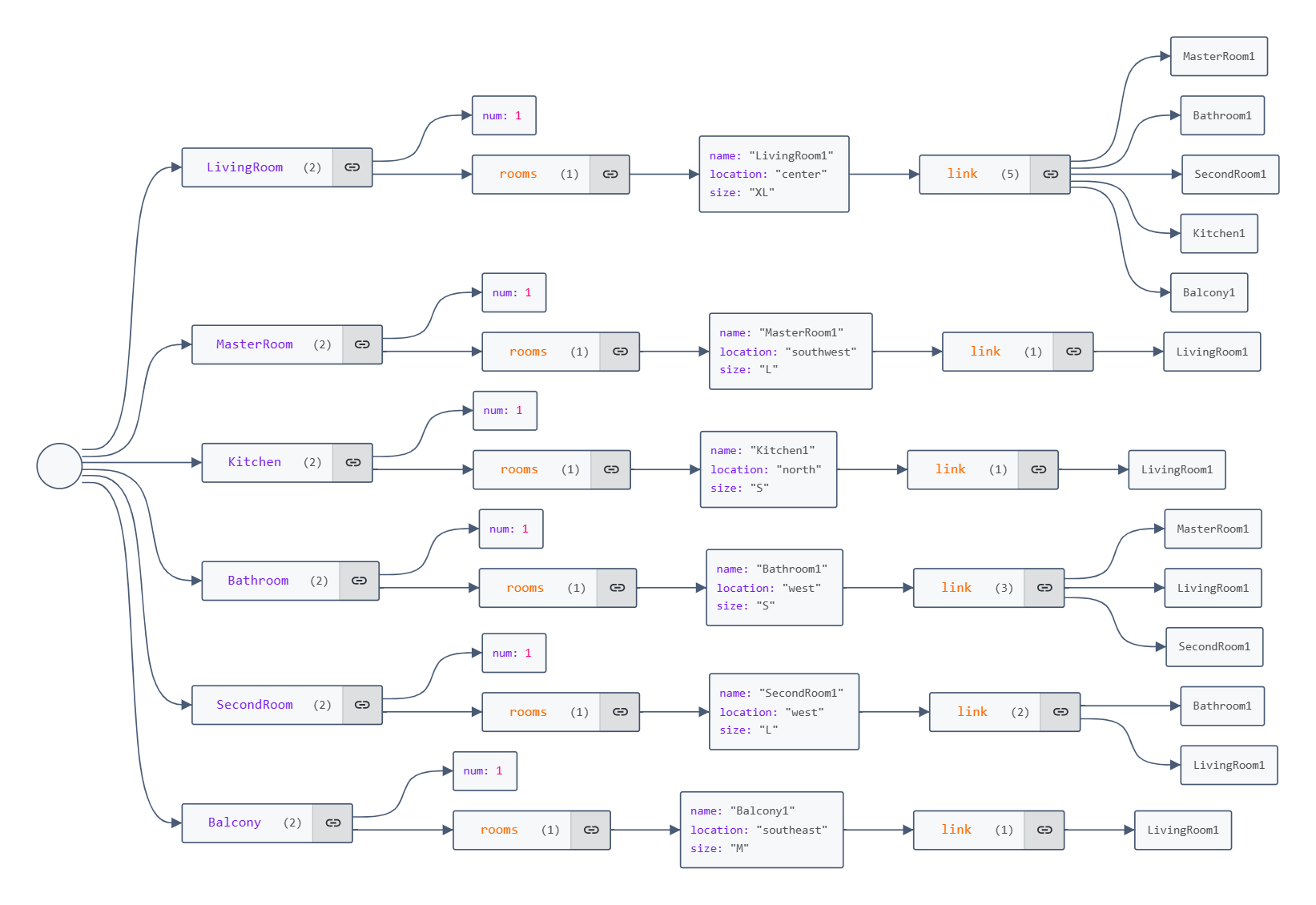}
    \caption{Input JSON}
    \label{fig:json}
\end{figure}

The information about the house, which  consists of the connection between rooms, the size of each room, and the location of each room, was analyzed based on the vector representations. Finally, the JSON files describing the house information were generated as another input(Figure \ref{fig:json}).

For evaluating generated floor plans against ground-truth data, we employ the following metrics \citep{leng-etal-2023-tell2design}. Micro IoU calculates the global Intersection over Union (IoU) by aggregating the intersections and unions of all room types in the floor plan. Macro IoU averages the IoU scores for different types of rooms, reflecting the model's accuracy per room type.
    \[
    \text{Micro-IoU} = \frac{\sum_{r=1}^{R} I_r}{\sum_{r=1}^{R} U_r},\quad\text{Macro-IoU} = \frac{1}{R} \sum_{r=1}^{R} \frac{I_r}{U_r}
    \]
where $I_r$ and $U_r$ represent the intersection and union areas of the ground-truth and predicted rooms for the $r$-th room type, respectively. $R$ is the total number of room types.
\subsection{LLM parsing}
 In our method, Llama3, gpt-4-turbo and moonshot-v1-8k (moonshot) are used to get the JSON-format prompt from linguistic prompts. The accuracy of extracting room information from text using three different LLM methods is compared, as shown in Table \ref{tab:llm-parsing}.

\begin{table}[ht]
\centering
\caption{Comparison of the accuracy of different LLMs in extracting room information}
\label{tab:llm-parsing}
\begin{tabular}{cccc}
\hline
LLM                     & Type & Size & Location \\
\hline
Llama3  & 90.10 & 33.48 & 43.62    \\
gpt-4-turbo  & 93.88  & 43.42 &53.90     \\
moonshot   & 93.81 & 48.44 &52.95     \\
\hline
\end{tabular}
\end{table}

From the table, it can be observed that moonshot and gpt-4-turbo perform comparably and outperform Llama3. Specifically, gpt-4-turbo excels in the recognition of type and location, though its performance in recognizing size is slightly inferior. Conversely, moonshot exhibits the best performance in recognizing size and demonstrates superior overall performance. Overall, LLMs accurately recognize room types, but need improvement in identifying size and location. This is primarily due to the broad definitions of size and location, as well as the imprecise textual descriptions. In practical design, this issue can be addressed through subsequent interactive editing.

\subsection{Floor plan generation}
\label{section:Results}
\begin{figure*}[ht]
    \centering
    \includegraphics[width=\textwidth]{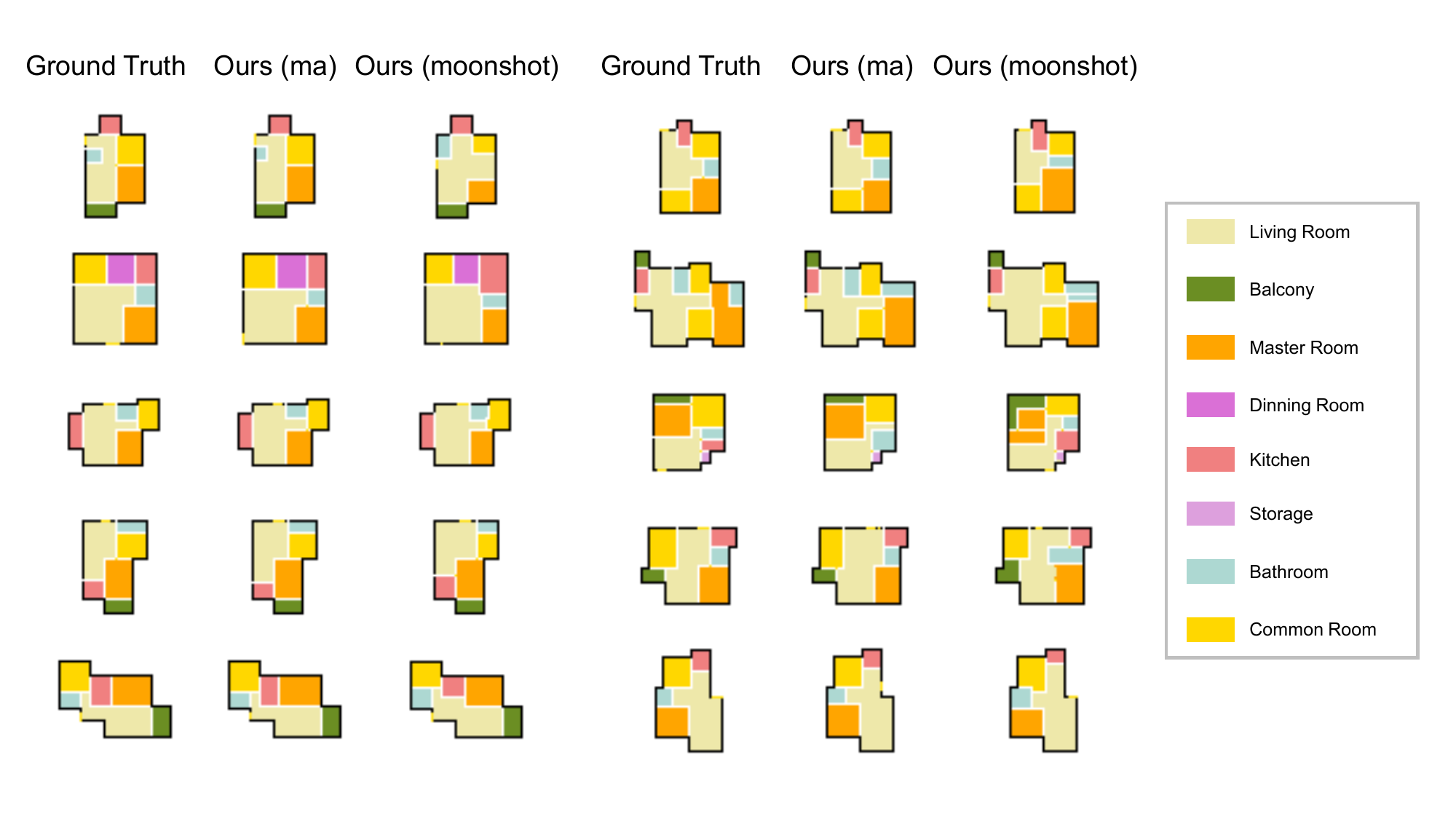}
    \caption{Cases of room plans generated by ChatHouseDiffusion}
    \label{fig:cases}
\end{figure*}
Employing T5 as the text encoder, our results were compared with those of Tell2Design \citep{leng-etal-2023-tell2design,liChatdesignBootstrappingGenerative2024}, utilizing linguistic prompts and contour boundaries in the test set to generate floor plans. Manual annotation (ma) refers to extracting data in JSON format directly from floor plans, which can generally be considered ground truth. This method is significantly more accurate compared to extracting data from text using LLMs.

\begin{table}[ht]
\centering
\caption{IoU scores between ground-truth and generated floor plans}
\label{tab:iou}
\begin{tabular}{ccc}
\hline
Method                      & Micro-IoU & Macro-IoU \\
\hline
Obj-GAN                     & 10.68     & 8.44      \\
CogView                     & 13.30     & 11.43     \\
Imagen                      & 12.17     & 14.96     \\
T2D (w/o boundary)      & 35.95     & 29.95     \\
T2D                      & 54.34     & 53.30     \\
ChatDesign             & 58.31     & 55.43     \\
ChatDesign-iterative         & 54.57     & 57.24     \\
ChatHouseDiffusion (Llama3)  & 54.64     & 51.22     \\
ChatHouseDiffusion (gpt-4-turbo)  & 60.51     & 56.27     \\
ChatHouseDiffusion (moonshot)   & \textbf{60.57}     & \textbf{57.34}     \\
ChatHouseDiffusion (ma) & \textbf{85.04}     & \textbf{82.32}     \\
\hline
\end{tabular}
\end{table}

\begin{figure}[ht]
    \centering
    \includegraphics[width=\linewidth]{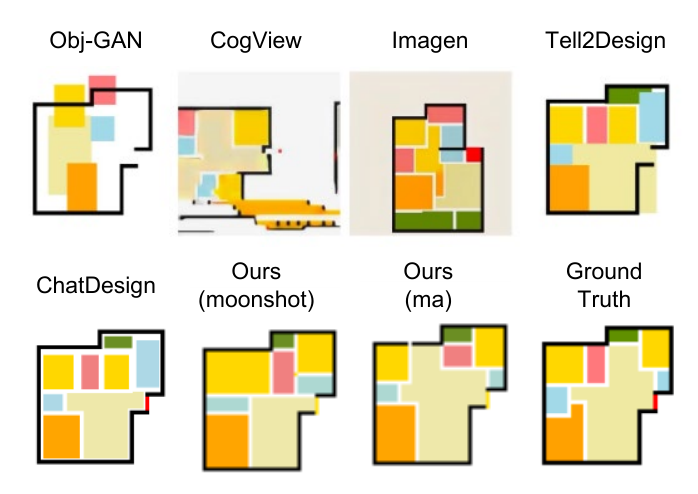}
    \caption{Comparison of generated room plans from different methods}
    \label{fig:case_compare}
\end{figure}

The IoU outcomes are as illustrated in Table \ref{tab:iou}. Figure \ref{fig:cases} presents the generated results of several floor plans, including those annotated manually (ma) and generated using the moonshot-v1-8k method (moonshot). Comparison of the generated results for different methods is performed using the prompt in the supplementary material, and the generated floor plans are illustrated in Figure \ref{fig:case_compare}.

\begin{figure*}[ht]
    \centering
    \includegraphics[width=0.95\textwidth]{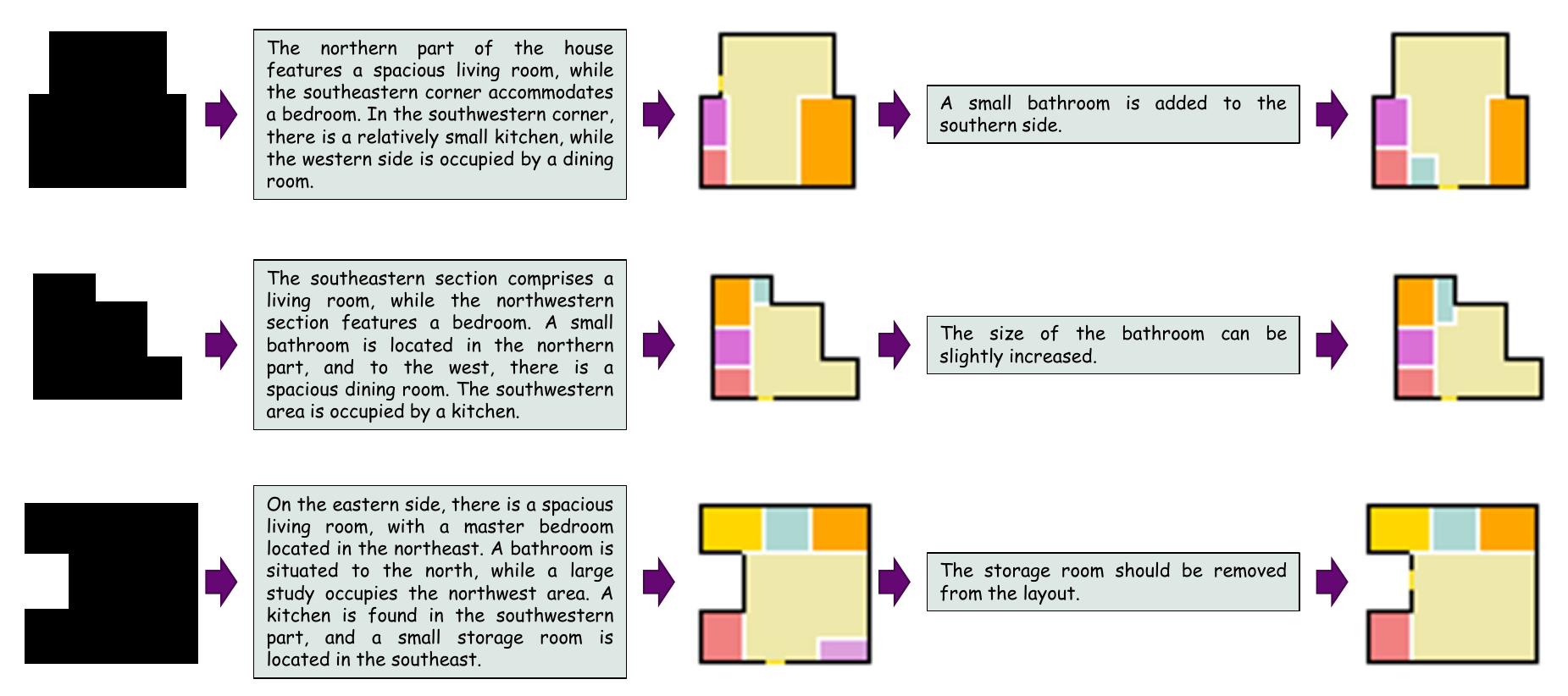}
    \caption{Cases of floor plan generation and editing}
    \label{fig:editing_results}
\end{figure*}
In comparison to previous methods, our approach has demonstrated significant improvements in terms of IoU scores. The IoU score achieved by ChatHouseDiffusion (manual annotation) is notably higher than other methods, highlighting the potential of our approach. This indicates that as long as the LLM accurately parses the text, better results can be achieved. Among the LLMs we have experimented with, ChatHouseDiffusion (moonshot) has exhibited the best performance, surpassing existing methods and further validating the effectiveness of our approach. According to Figure \ref{fig:cases}, it can be observed that the floor plans generated using our method are generally consistent with the ground truth. Minor local differences may exist but can be adjusted through subsequent edits. A comparison between our method and other methods, as shown in Figure \ref{fig:case_compare}, reveals that the floor plans generated by our method perfectly conform to the outer contour requirements and exhibit internal layouts that are closer to the ground truth and more reasonable.

\subsection{Floor plan editing}
For editing floor plans, a series of case studies were conducted to demonstrate the effects of room additions, modifications, and deletions. The results, as depicted in Figure \ref{fig:editing_results}, illustrate that our method can accurately achieve localized room editing without altering the overall floor plan.

\subsection{Discussion}
\label{section:Discussion}
\subsubsection{Generation}

During the generation phase, textual descriptions and room contour information were fully utilized in our method. ChatHouseDiffusion (ma) employed entirely accurate descriptive information, achieving both Micro-IoU and Macro-IoU values exceeding 80\%, making further improvements challenging with the limited information available. Specific cases reveal that differences between the generated results and the ground truth include room shapes and the relative positions of rooms, suggesting that incorporating these features could yield more precise generation.

Comparing the performance of various LLMs, moonshot-v1-8k and gpt-4-turbo achieved very similar results, while Llama3 performed poorly. Both moonshot-v1-8k and gpt-4-turbo have essentially reached the current upper limit of LLMs' parsing capabilities, but significant differences from the ground truth remain. Two reasons can be summarized: first, the Tell2Design dataset contains some imprecise textual descriptions that do not accurately reflect the actual floor plan; second, textual descriptions significantly impact IoU. For example, for smaller rooms, the locations might be very close, yet the IoU could be zero, and such situations can greatly affect the final results. Therefore, in practical applications, it is crucial to use accurate descriptions or edit imprecise descriptions to ensure the accuracy of the final results.

\subsubsection{Editing}

In the editing phase, the addition, deletion, and modification of rooms can be effectively implemented. In some cases, issues with modifying other rooms still exist, which can be addressed by adjusting the threshold $\mathrm{\tau}$. However, due to the inherent limitations of language expression, precise adjustments, such as the exact size of rooms, cannot be achieved. Future improvements could integrate a graphical interface that utilizes drag-and-drop features to enhance the interactivity of floor plan editing, resulting in a more practical floor plan interaction.

\section{Conclusion}
\label{section:Conclusion}
In this study, we have presented ChatHouseDiffusion, an advanced diffusion model tailored for the automated generation and editing of floor plans using text prompts. This model significantly enhances the architectural design process by incorporating classifier-free guidance diffusion and masks, coupled with LLM and graphormer to interpret and utilize textual descriptions effectively.

Our approach has demonstrated superior performance in generating floor plans that closely adhere to specified design requirements, as evidenced by the high IoU scores when compared with previous methods. Furthermore, our editing methodology, which is based on localized adjustments using a cross-attention map, has been shown to adeptly handle specific design changes without necessitating a complete redesign. This capability is crucial for practical architectural design applications.

Overall, the results of our experiments and the feedback from case studies suggest that ChatHouseDiffusion pushes the boundaries of automated floor plan design, paving the way for more intuitive and efficient design processes in the future.

\section{Supplementary material}
\subsection{Data preprocess}
\label{section:Data preprocess}
\begin{figure}[ht]
    \centering
    \begin{minipage}[t]{0.4\linewidth}
        \centering
        \includegraphics[width=0.6\textwidth]{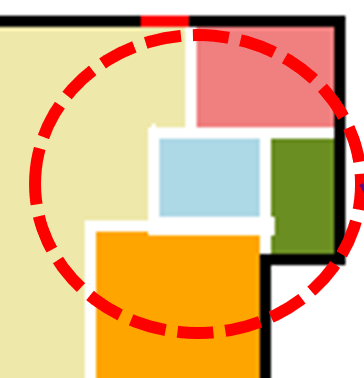}
        \caption{Dislocation}
        \label{fig:dislocation}
    \end{minipage}
    \begin{minipage}[t]{0.28\linewidth}
        \centering
        \includegraphics[width=\textwidth]{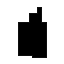}
        \caption{Input}
        \label{fig:boundary}
    \end{minipage}
    \begin{minipage}[t]{0.3\linewidth}
        \centering
        \includegraphics[width=\textwidth]{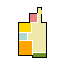}
        \caption{Output}
         \label{fig:output}
    \end{minipage}
\end{figure}

The original dataset includes 80,788 images, and the resolution of each image is 256 × 256. Some of the original data set has pixel dislocation problem as shown in Figure \ref{fig:dislocation}. Moreover, resizing the original image directly will aggravate this problem. To redraw the floor plan, the pixel semantic information needs to be extracted from the original data. Then using the coordinates of key points to describe the floor plan. So, the vector representation of the floor plan was created based on the original image for each sample. Then the house boundaries (Figure \ref{fig:boundary}) and floor plans (Figure \ref{fig:output}) were extracted as a resolution of 64 × 64. 

\begin{figure}[ht]
        \centering
        \includegraphics[width=0.6\linewidth]{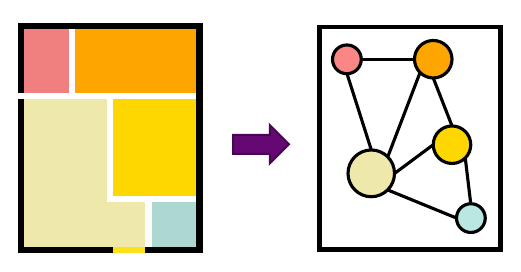}
        \caption{Connection analysis }
         \label{fig:connection}
\end{figure}

\begin{figure}[ht]
    \centering
    \includegraphics[width=0.6\linewidth]{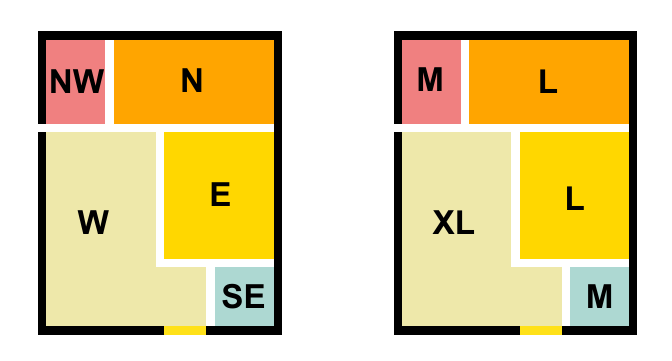}
    \caption{Size and location}
     \label{fig:size}
\end{figure}

To analyze the house information, the doors inside the house were identified firstly, and then the connection analysis (Figure \ref{fig:connection}) was conducted based on the locations of the interior doors. Besides, the area of each room was calculated. The size and location were assigned to each room for JSON generation (Figure \ref{fig:size}). 

\subsection{Prompt design}
\label{section:Prompt design}
The LLM is accessed via the OpenAI API, allowing for easy integration with other LLMs using API keys.

When building the prompt, the user's text and the output schema are combined. The output schema is shown below. The prompt contains the meaning of each attribute in JSON and the legal type.

\begin{lstlisting}[caption={Output Schema}, label={lst:json}, aboveskip=5pt, belowskip=-3pt]
{
  "properties": {
    "rooms": {
      "title": "Rooms",
      "type": "array",
      "items": {"$ref": "#/definitions/Room"}
    }
  },
  "required": ["rooms"],
  "definitions": {
    "RoomType": {
      "title": "RoomType",
      "description": "An enumeration.",
      "enum": ["LivingRoom", "MasterRoom", "Kitchen", "Bathroom", "DiningRoom", "CommonRoom", "SecondRoom", "ChildRoom", "StudyRoom", "GuestRoom", "Balcony", "Entrance", "Storage"]
    },
    "LocationType": {
      "title": "LocationType",
      "description": "An enumeration.",
      "enum": ["north", "northwest", "west", "southwest", "south", "southeast", "east", "northeast", "center"]
    },
    "SizeType": {
      "title": "SizeType",
      "description": "An enumeration.",
      "enum": ["XL", "L", "M", "S", "XS"]
    },
    "Room": {
      "title": "Room",
      "type": "object",
      "properties": {
          "name": {
            "title": "Name",
            "description": "The name of the room. Ensure it is unique.",
            "type": "string"
          },
          "type": {
            "description": "The type of the room.",
            "allOf": [{"$ref": "#/definitions/RoomType"}]
          },
          "link": {
            "title": "Link",
            "description": "The names of the rooms this room is connected to.",
            "type": "array",
            "items": {"type": "string"}
          },
          "location": {
            "description": "The location of the room within the layout. Top represents the north, bottom represents the south.",
            "allOf": [{"$ref": "#/definitions/LocationType"}]
          },
          "size": {
            "description": "The size of the room, calculated as a proportion of the entire layout outline.",
            "allOf": [{"$ref": "#/definitions/SizeType"}]
          }
      },
      "required": ["name", "link"]
    }
  }
}
\end{lstlisting}
When implementing the editing function, the last generated result will be added to the prompt to prompt the LLM to edit.

\subsection{UI design}
\label{section:UI design}
\begin{figure*}[ht]
    \centering
    \includegraphics[width=\linewidth]{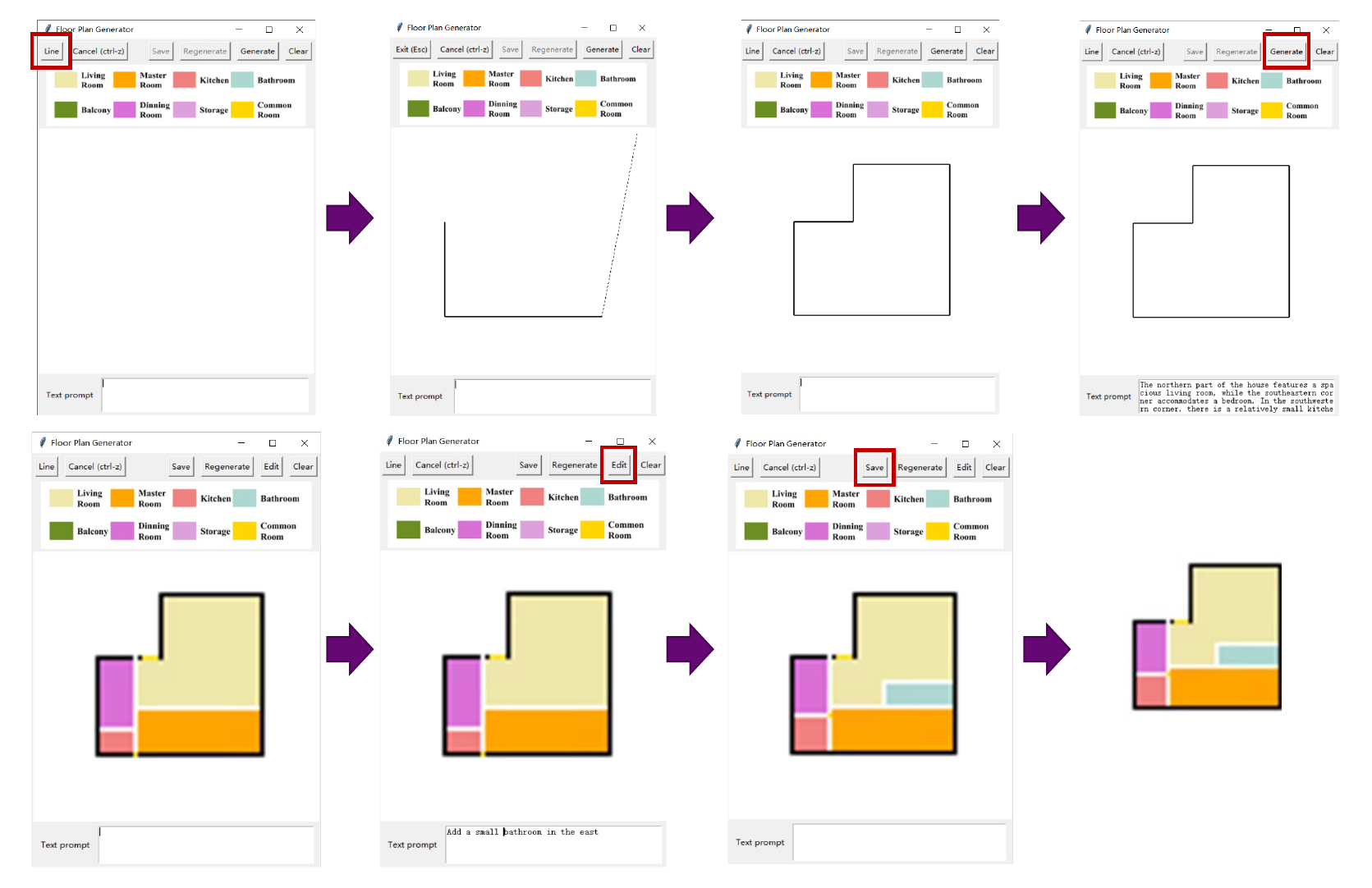}
    \caption{UI design of ChatHouseDiffusion}
    \label{fig:ui}
\end{figure*}

A user interface (UI) design was also implemented using the tkinter library in Python, enabling the drawing of outlines, inputting text prompts, and generating and editing floor plans. This is illustrated in Figure \ref{fig:ui}.

To facilitate user operations, we have incorporated numerous useful features. During the drawing phase, these include features such as dashed line indication, orthogonal snap, endpoint snap, and the ability to undo the previous drawing step. In the generation and editing phase, features include remembering the last input text and regenerate floor plan by modifying the seed. Users can refer to our GitHub repository instructions (to be open-sourced in the future) to set up their own API token and implement floor plan generation and editing.

\subsection{Prompt for Figure \ref{fig:case_compare}}
\label{section:Case study}
The north side of this home is not complete without the balcony. Access to the approximately 16 sq ft area can be made through the living room or through the common room beside it. Bathroom 1 is in the eastern section of the home. It is located next to the living room and is approximately 15 sq ft. The larger of the two, Bathroom 2, is approximately 30 sq ft. It is between the master bedroom and common area 2, along the western side of the house. Common room 1 occupies the northeast corner of the property. At roughly 80 sq ft it is conveniently located next to the balcony. Common room 2 is nearly 100 sq ft. Occupying the northwest corner, it is easily accessible from the kitchen beside it, or the shared access from the living area. The kitchen is positioned on the north side of the house, between the living room and second common area. It measures about 50 sq ft. The living room is conveniently located in the southeast corner of the home. It spans approximately 250 sq ft while offering access to almost every room in the house. Located in the southwest corner of the home is the master bedroom. This space is approximately 120 sq ft and is positioned next to the living room. 

\bibliography{ref}

\end{document}